# Quantum-Secure Hybrid Blockchain System for DID-based Verifiable Random Function with NTRU Linkable Ring Signature

Bong Gon Kim[1], Dennis Wong[2], and Yoon Seok Yang[3]

[1]Department of Computer Science, Stony Brook University, New York, USA,

[2]Faculty of Applied Science, Macao Polytechnic University, Macao, China,

[3]Department of Computer Science, SUNY Korea University, South Korea

**ABSTRACT**

*In this study, we present a secure smart contract-based Verifiable Random Function (VRF) model, addressing the shortcomings of existing systems. As quantum computing emerges, conventional public key cryptography faces potential vulnerabilities. To enhance our VRF's robustness, we employ post-quantum Ring-LWE encryption for generating pseudo-random sequences and a NTRU lattice-based linkable ring signature scheme. Given the computational intensity of this approach and associated on-chain gas costs, we propose a hybrid architecture of VRF system where both on-chain and off-chain can communicate in a scalable and secure way. Our decentralized VRF employs multi-party computation (MPC) with blockchain-based decentralized identifiers (DID), ensuring the collective efforts of enhanced randomness and security. We show the security and privacy advantages of our proposed VRF model with the approximated estimation of overall temporal and spatial complexities. We also evaluate our VRF MPC model's entropy and outline its Solidity smart contract integration. This research also provides a method to produce and verify the VRF output's proof, optimal for scenarios necessitating randomness and validation. Lastly, using NIST SP800-22 test suite for randomness, we demonstrate the commendable result with a 97.73% overall pass rate on 11 standard tests and 0.5459 of average p-value for the total 176 tests.*

## 1 Introduction

Cryptographic protocols, especially in blockchain systems, fundamentally rely on randomness. The Verifiable Random Function (VRF) [1] is a primary cryptographic utility that produces verifiable pseudo-random numbers with uniqueness, uniformly distributed randomness property, and proof of correctness. Applied in areas ranging from lotteries to proof-of-stake [2] and privacy-preserving protocols, a genuine challenge is developing VRFs that truly emulate randomness. For instance, Algorand [3] utilizes VRFs for its cryptographic self-selection, orchestrating the formation of Selected Verifiers (SV) and directing block generation leadership, underscoring VRF's central role in decentralized integrity.

The evolving cryptographic landscape underscores the need for VRFs to adapt. Contemporary cryptographic models, despite their sophistication, face potential vulnerabilities, accentuated by the advent of quantum computation. Specifically, Shor's algorithm [4], a potent quantum formula, threatens the resilience of classical public key cryptographic methodologies. Recognizing the quantum revolution, there's a pressing call to incorporate post-quantum cryptographic innovations. These progressive shifts prioritize not only quantum resistance but also cryptographic efficacy.

As such, the lattice-based cryptographic schemes, which root back to Ajtai's worst-case to average-case reduction [5], provide a compelling cryptographic foundation. The Learning With Errors (LWE) problem [6] and its ring-based counterpart, Ring-LWE [7], and Small Integer Solution (SIS) problem (average-case problem in [5]), and Shortest Vector Problem [8] (the most famous computational problem on lattices), have gained prominence for their inherent resilience against quantum attacks [9]. The Ring-LWE variant introduces structured noise and structured secrets over polynomial rings, offering efficiency gains while retaining the security attributes [10].

Ring-LWE distinguishes itself from traditional lattice-based approaches by optimizing computations and minimizing key sizes, harnessing the algebraic structure of number theoretic rings. In the post-quantum landscape, where conventional cryptographic constructs falter under quantum algorithms like Shor's [4], Ring-LWE encryption, with its inherent resistance to quantum decryption attempts, ensures that encrypted data remains secure even in the face of quantum computational capabilities.

NTRU, introduced in 1996 [11] by Hoffstein, Pipher, and Silverman, represents a significant advancement in cryptography, leveraging the complexity of lattice problems such as R-SIS and R-LWE [12]. Its quantum resistance has positioned NTRU in the NIST Post-Quantum Cryptography Standardization project [13], marking its relevance in the ongoing transition to quantum-secure communications.

In the specific context of the decentralized systems like Ethereum, the VRF is also crucial for consensus algorithms such as validator selections, block validations, and attestations. The unpredictable randomness demanded by such platforms can be securely achieved using the Ring-LWE encryption with NTRU linkable ring signature. For instance, when the parties collaborate to generate a random output collectively in a decentralized environment, the Ring-LWE encryption and NTRU can provide high randomness and verifiablility, while resistant to external (including quantum) threats.

However, the Ring-LWE encryption introduces considerable computational overheads. For blockchain ecosystems, the integration of Ring-LWE can potentially inflate transactional costs such as *gas costs*. To alleviate these computational and cost impediments, off-chain computation emerges as an optimal solution. Within the scope of MPC-based VRFs, off-chain computation allows for the intricate cryptographic operations to be executed outside the primary blockchain. This not only conserves on-chain resources but significantly reduces associated gas costs.

To provide the off-chain computations with authenticity and validity, we propose the Decentralized Identifiers (DID) [14]-based linkable ring signature scheme [15] on NTRU lattice. The signature scheme offers the integrity for the off-chain computation with delegated key generation from on-chain to off-chain via a key encapsulation mechanism (KEM) [16]. Thus, they serve as a robust mechanism to attest to the correctness of off-chain computations, ensuring that they adhere to expected behavior and outputs. Through the DID-based signature scheme with NIZK proof, the blockchain network can trust the authenticity and accuracy of data derived off-chain, bolstering the overall trustworthiness of decentralized systems. Our main contributions to this paper are:

– Proposing a post-quantum VRF leveraging DIDs and the linkable ring signature scheme on NTRU lattice, spotlighting off-chain validation.
– Constructing a smart contract-based VRF using MPC and Ring-LWE, as inspired by Clercq et al.'s research [17].
– Illustrating our Solidity and Python implementations, algorithmic procedures, and providing complexity estimates (Fig. 5).
– Delivering security proofs, privacy analyses, and Ring-LWE-based VRF entropy assessments.
– Presenting NIST SP800-22 results, showcasing a 97.73% pass rate across 11 tests and a 0.5459 average $p$-value from 176 tests (Fig. 2, 3, 4 and Table 1).

The structure of this paper is presented as follows: Section 2 delves into the intricacies of VRF and Ring-LWE encryption, establishing foundational context. In Section 3, we outline our novel VRF model, detailing its formal instantiation with NIZK proof, its decentralized algorithmic characteristics, and integrations with smart contracts. Section 4 focuses on the security parameters, emphasizing the post-quantum adversarial advantage probability and essential VRF key verifications. Also, Section 5 scrutinizes temporal and spatial complexities. In Section 6, we present in-depth entropy assessment of our MPC-based VRF, exploring the theoretical bounds of its randomness. Also, we summarize the experimental outcomes from the NIST SP800-22 [18] test suite, consisting of 11 test cases for randomness evaluation, and illustrate our implementation using Solidity and Ganache. Finally, we consolidate our discussions and findings in Section 7.

## 2 Preliminaries

### 2.1 VRF Constructions

Verifiable Random Functions (VRF) serve as the backbone for various cryptographic protocols where both unpredictability and verifiability are required. These functions are akin to Pseudo-Random Functions (PRFs) but come with an added property: they can generate a proof for each output, ensuring the verifier that a particular random number was produced correctly [1]. Typically, a VRF consists of three polynomial-time algorithms:

– **KeyGen**: which outputs a public key *PK* and a secret key *SK*.
– **Evaluate**: using *SK* and input *x*, it produces an output *y* and a proof $\pi$.
– **Verify**: with *PK*, *x*, *y*, and $\pi$, it ensures the validity of *y*.

The original conception of VRF by Micali et al. [1] pivoted on an RSA-based verifiable unpredictable signature scheme, leveraging randomness via the Goldreich-Levin hardcore bit construction. Subsequent innovations, such as the number-theoretic exponentiation-based PRF by Naor and Reingold [19], and theoretical advancements by Joux et al. showing that a Computational Diffie-Hellman (CDH) problem remain as hard as the discrete logarithm (DL) problem, while the corresponding Decisional Diffie-Hellman (DDH) problem becomes easy with certain multiplicative groups [20], further enriched the VRF landscape. The discovery of the group is based on the bilinear pairing methods [21] and led to novel signature-based VRFs like Lysyanskaya's [22], which capitalized on prior works and the notion of an admissible hash function (AHF). The rigorous demands for cryptographic strength in hash functions [23, 24] steered research towards constructing AHFs that hold firm in standard security models [25]. Recently, noteworthy advancements in VRFs have been made by Tibor and colleagues, innovating AHF schemes that fortify the instantiation of VRF with large input size and full adaptive security [26–29].

In the blockchain framework, the VRF becomes more pronounced. Their deterministic yet unpredictable nature is pivotal in consensus algorithms, ensuring fairness and thwarting adversaries from biasing outcomes. To enhance the fairness, collective efforts for randomness are proposed using ChaCha20 stream cipher in [30] and game-theoretic principles [31]. Kim et al. also utilized the de Bruijn sequence [32] with balanced property for VRF construction within blockchain ecosystems.

However, the looming quantum computing era casts a shadow on these classical cryptographic systems. Given that VRFs often lean on problems that could be susceptible to quantum algorithms, such as the DL problem or large prime numbers factorization, there is an imperative to future-proof these constructs.

Our proposed approach diverges from the conventional methods, by incorporating a blockchain-based VRF via Multiparty Computation (MPC) with Ring-LWE encryption and NTRU linkable ring signature. Our VRF scheme utilizes the NTRU lattice cryptography to generate the DID-based linkable ring signature with VRF proof, which differentiates from the recent post-quantum VRF instantiaions of XMSS signature-based X-VRF scheme [33] or conventional ring signature-based VRF scheme [34]. To this end, we aim to provide enhanced post-quantum security, scalability, and efficiency in random value generation for decentralized systems.

## 2.2 Lattice-Based Cryptography

Lattice-based cryptography, heralded as a beacon for post-quantum cryptographic systems, pivots on the intricate mathematical structures of lattices. This burgeoning field offers dual benefits: adaptability across varied cryptographic applications and perceived quantum-resilient security.

A *lattice L* is formally characterized as a discrete subgroup of $\mathbb{R}^n$. It can be thought of as the grid generated from taking all integral linear combinations of a set of $n$ linearly independent vectors $\mathbf{b}_1, \mathbf{b}_2, \ldots, \mathbf{b}_n$ in $\mathbb{R}^n$. These vectors form a *basis* for the lattice. Symbolically,

$$L(\mathbf{b}_1, \mathbf{b}_2, \ldots, \mathbf{b}_n) = \{\sum_{i=1}^{n} x_i \mathbf{b}_i : x_i \in \mathbb{Z}\}$$

From a cryptographic viewpoint, the intractability of certain lattice problems, notably the Shortest Vector Problem (SVP) and the Learning With Errors (LWE) problem, formulates the bedrock of the lattice-based cryptography's security [5, 35].

## 2.3 NTRU Lattice Crptography

Originally presented at CRYPTO '96 and further detailed in 1998, NTRU's efficiency and security against quantum attacks have made it a noteworthy candidate for post-quantum cryptography. The algorithm underwent standardization processes, like IEEE P1363.1, and faced cryptanalysis leading to parameter adjustments. The U.S. patent for the original NTRU system (U.S. patent 6081597) was released to the public in March 2017 (expired in August 2017) and the various past refinements are referred to [36].

**NTRU Lattice** Let $\phi \in \mathbb{Z}[x]$ be a monic polynomial (i.e., cyclotomic polynomial), and $q$ be a positive integer. The NTRU lattice $\Lambda$ is constructed using four polynomials $f, g, F$, and $G$ in the quotient ring $\mathbb{Z}[x]/(\phi)$. These polynomials form the secret key basis and satisfy the following congruence, which forms the NTRU equation [37]:

$$fG - gF = q \mod \phi \tag{1}$$

Provided $f$ is invertible modulo $q$, we can define the public key polynomial $h$ as:

$$h \leftarrow g \cdot f^{-1} \mod q \tag{2}$$

Detailed algebraic proof of how these equations play the NTRU basis operation is referred to the Theorem 2 in Appendix A of [38].

**Gram-Schmidt Orthogonalization (GSO)** The Gram-Schmidt process is a method for orthogonalizing a set of vectors in an inner product space, which is commonly the Euclidean space $\mathbb{R}^n$. Given a basis for a lattice given by the vectors $\mathbf{b}_1, \mathbf{b}_2, \ldots, \mathbf{b}_n \in \mathbb{R}^n$, the Gram-Schmidt process converts this basis into an orthogonal basis $\mathbf{b}_1^*, \mathbf{b}_2^*, \ldots, \mathbf{b}_n^*$, where each $\mathbf{b}_i^*$ is orthogonal to all previous vectors $\mathbf{b}_1^*, \ldots, \mathbf{b}_{i-1}^*$. The process works as follows in Algorithm 1:

The set of orthogonal vectors $\mathbf{b}_1^*, \ldots, \mathbf{b}_n^*$ span the same subspace as the original vectors $\mathbf{b}_1, \ldots, \mathbf{b}_n$. These orthogonal vectors can be easier to work with, especially when performing tasks like finding shortest vectors in the lattice (the LLL algorithm [39], for instance, uses Gram-Schmidt orthogonalization as a subroutine). The faster GSO computing algorithm [12] by Lyubashevsky et al. has made NTRU lattice more practical along with their Compact Gaussian Sampling (CGS) [12], which is used in this paper's NTRU-based linkable ring signature scheme in Algorithm 4.

---

**Algorithm 1** Gram-Schmidt Orthogonalization for NTRU lattice
---
**Require:** A set of linearly independent vectors $\mathbf{b}_1, \mathbf{b}_2, \ldots, \mathbf{b}_n \in \mathbb{R}^n$
**Ensure:** An orthogonal set of vectors $\mathbf{b}_1^*, \mathbf{b}_2^*, \ldots, \mathbf{b}_n^*$
1:  $\mathbf{b}_1^* \leftarrow \mathbf{b}_1$
2:  **for** $i \leftarrow 2$ to $n$ **do**
3:      $\mathbf{b}_i^* \leftarrow \mathbf{b}_i$
4:      **for** $j \leftarrow 1$ to $i-1$ **do**
5:          $\text{proj}_{\mathbf{b}_j^*}(\mathbf{b}_i) \leftarrow \left( \frac{\langle \mathbf{b}_i, \mathbf{b}_j^* \rangle}{\langle \mathbf{b}_j^*, \mathbf{b}_j^* \rangle} \right) \mathbf{b}_j^*$
6:          $\mathbf{b}_i^* \leftarrow \mathbf{b}_i^* - \text{proj}_{\mathbf{b}_j^*}(\mathbf{b}_i)$
7:      **end for**
8:  **end for**

---

## 2.4 NTRU-based Signature Generation

The general NTRU-based digital signature scheme follows the Gentry-Peikert-Vaikuntanathan (GPV) framework [40] for lattice-based signatures. The scheme's operation can be summarized as follows:

**Key Generation**

1. **Public Key**: A full-rank matrix $A \in \mathbb{Z}_q^{n \times m}$ is generated, where $m > n$. This matrix $A$ generates a $q$-ary lattice $\Lambda$.
2. **Private Key**: A matrix $B \in \mathbb{Z}_q^{m \times m}$ is generated, which forms a basis for the lattice $\Lambda^\perp$ through the GSO process in Algorithm 1, orthogonal to $\Lambda$ modulo $q$. This means for any vector $x \in \Lambda$ and any vector $y \in \Lambda^\perp$, their dot product satisfies $\langle x, y \rangle = 0 \mod q$.

**Signature Generation**

1. **Hash Function**: A hash function $H$ maps the message $m$ to a hash value in $\mathbb{Z}_q^n$.
2. **Initial Signature Vector**: An initial signature vector $c_0 \in \mathbb{Z}_q^m$ is computed such that $c_0$ has a preimage under $A$ and $c_0 A^t = H(m)$, where $A^t$ is the transpose of $A$.
3. **Close Vector Calculation**: The private key matrix $B$ is used to compute a vector $v$ in $\Lambda^\perp$ that is close to $c_0$.
4. **Final Signature**: The signature $s$ is the difference between the initial signature vector $c_0$ and the close vector $v$, i.e., $s = c_0 - v$. The signature $s$ is valid if $sA^t = c_0 A^t - vA^t = H(m) - 0 = H(m)$, given that $v \in \Lambda^\perp$ and $vA^t = 0$.

**Verification** To verify the signature, one needs to check two things:

1. $s$ is indeed a "short" vector in $\mathbb{Z}_q^m$, meaning its entries are small integers, which is a property required for the security of the scheme.
2. $s$ satisfies $sA^t = H(m)$. If both conditions hold, the signature is considered valid.

## 2.5 Ring-LWE Cryptography

Ring-LWE (Ring Learning With Errors) cryptography is based on the Ring-LWE problem's hardness, adapted from the Learning With Errors (LWE) problem to polynomial rings. It is designed to be secure against quantum attacks and efficient for practical use. Typically, Ring-LWE operates over a ring $R = \mathbb{Z}[x]/(f(x))$, with $f(x)$ being a monic irreducible polynomial, typically $f(x) = x^n + 1$ for some power-of-two $n$, and $q$ a prime number. The operations in Ring-LWE are modulo $f(x)$ and $q$.

**Encryption and Decryption**

– **Key Generation:** A secret key is a small polynomial $s(x) \in R_q$. The public key is $(a(x), b(x) = a(x) \cdot s(x) + e(x))$, with $a(x)$ random in $R_q$ and $e(x)$ a small error polynomial.
– **Encryption:** To encrypt a message $m(x)$, compute the ciphertext $(c_1(x), c_2(x))$ where:

$$c_1(x) = a(x) \cdot r(x) + e'(x)$$
$$c_2(x) = b(x) \cdot r(x) + m(x)$$

Here, $r(x)$ is random and small, and $e'(x)$ is another small error polynomial.
– **Decryption:** Decrypt by computing:

$$c_2(x) - s(x) \cdot c_1(x) = m(x) + r(x) \cdot e(x) - s(x) \cdot e'(x)$$

The message $m(x)$ is recovered by correcting the small error term.

**VRF Instantiability** The Ring-LWE encryption's security relies on the computational infeasibility of solving the Ring-LWE problem. Its efficiency benefits from polynomial ring structures that allow fast multiplication, such as the Number Theoretic Transform (NTT). This Ring-LWE's computational efficiency makes it an ideal choice for cryptographic tasks such as pseudo-random number generation and Verifiable Random Functions (VRF). Explorations merging Ring-LWE with random number generation have been notably impactful. For instance, Abraham's study delineates a robust VRF scheme powered by Ring-LWE, underscoring the dual benefits of speed and security [41].

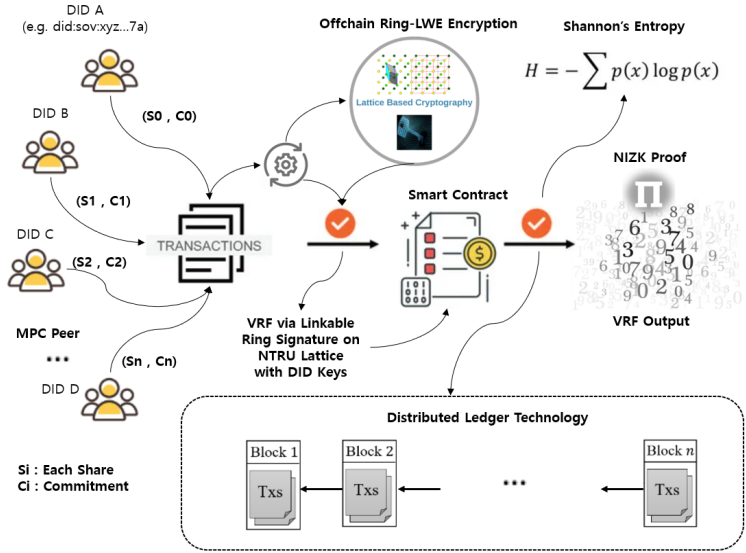

**Fig. 1.** A quantum-secure hybrid system for DID-based verifiable random function incorporating MPC and DIDs, where off-chain Ring-LWE encryption is executed with the linkable ring signature on NTRU lattice scheme via DKG.

## 3  Proposed VRF System

### 3.1  System Architecture

We introduce a novel VRF mechanism crafted for blockchain systems, leveraging Ring-LWE encryption [17] with the DID-based linkable ring signature scheme [15] on NTRU lattice, employing

**Algorithm 2** The main computation process that involves quantum-secure linkable ring signature on NTRU lattice, generating the VRF output and proof using the commitments of the participants. The computation is done using multi-party computation that ensures the privacy of the participants' shares and the correctness of the result.

**Require:** $roundId \in \mathbb{Z}, commitment \in \{0, 1\}^n$, where $n$ denotes 256-bits.
**Ensure:** None
1: **Require:** the $combinedBlockHash$ of the 3 most recent blocks $> 0$
2: $resultBytes \leftarrow$ byte array of length 32
3: $weightedCommitmentSum \leftarrow 0$
4: $shareSum \leftarrow 0$
5: **for** $i \leftarrow 0$ to $|participants[roundId]| - 1$ **do**
6:     $participant \leftarrow participants[roundId][i].addr$
7:     $share \leftarrow shares[roundId][participant]$
8:     $commitment \leftarrow commitments[roundId][participant]$
9:     $commitmentHash \leftarrow$ keccak256($commitment$)
10:     $weightedCommitment \leftarrow (commitmentHash \cdot share)$ mod n
11:     $weightedCommitmentSum \leftarrow (weightedCommitmentSum$
        $+weightedCommitment)$ mod n
12:     $shareSum \leftarrow (shareSum + share)$ mod n
13: **end for**
14: **if** $shareSum \neq 0$ **then**
15:     $mpcResult \leftarrow \lfloor weightedCommitmentSum/shareSum \rfloor$
16: **else**
17:     $mpcResult \leftarrow 0$
18: **end if**
19: $onchainSeed \leftarrow abi.encodePacked(block.number, block.timestamp, msg.sender,$
    $combinedBlockHash, mpcResult)$
20: $(c, K) \leftarrow Encaps(params, pk_{off})$     ▷ DKG Key Encapsulation
21: **Emit:** OnchainMpcSeedReady($onchainSeed, c, params, msg.sender$)
22: **Offchain:**
23: Cipher texts $(c1, c2) \leftarrow$ RLWE_PROCESSING($seed$)
24: $VRF_{\text{output}} \leftarrow$ keccak256($c1, c2$)
25: $K \leftarrow Decaps(params, c, sk_{off})$     ▷ DKG Key Decapsulation
26: $sk'_i \leftarrow$ DECRYPT($K, sk_{enc}$)
27: signature $\sigma \leftarrow$ NTRU_LinkableRingSign($sk', seed, VRF_{output}, R$)
    ▷ $R$ is the set of DID public keys
28: proof $\pi \leftarrow (VRF_{\text{output}}, Seed, \sigma)$
29: **Off-chain to On-chain:** SUBMIT_TO_BLOCKCHAIN($\pi$)
30: **On-chain:** VERIFYSIGN($\pi, R$)
31: **Emit:** COMPUTATIONFINISHED($roundId, \pi$)

a delegated key generation method with KEM [16]. We utilize lattice-based cryptography to generate random numbers from on-chain seed values in an off-chain environment, where the on-chain MPC participants collectively contribute to the seed values. The proposed hybrid VRF system is depicted in Fig. 1. Section 4 and 6.1 delve into a theoretical assessment of the security and entropy aspects of our VRF scheme.

In Fig. 1, we present a high-level architectural overview of our hybrid VRF system. The process begins with a distributed generation of the MPC-based seed on a smart contract. Each participant contributes a share and an associated commitment to generate this seed in Solidity. Upon the on-chain MPC seed-ready event, an off-chain blockchain listener, utilizing the Web3 protocol, is invoked and calls the function `RLWE_enc2()` to perform Ring-LWE encryption using the seed. The resultant ciphertexts $c1$ and $c2$ are dispatched to the smart contract via the `submitRLWEResult` function.

Our contribution is to ensure the integrity and validation of off-chain VRF output from the Ring-LWE encryption. This led to the adoption of a DID-based linkable ring signature scheme inspired by Tang et al [15]. Nonetheless, an issue arose due to the absence of an off-chain ring group $R$, whereas the $R$ exists in the on-chain that orchestrates MPC-based seed generation.

The issue was resolved through a delegated key generation (DKG) protocol with KEM [16], as shown in Algorithm 5. As the group forms for the MPC and ring signature operations on-chain, each participant, denoted as $i$, produces two key pairs: $(sk_i, pk_i)$ and $(sk'_i, pk'_i)$. While $(sk_i, pk_i)$ is tailored for on-chain tasks, $(sk'_i, pk'_i)$ is designated as the "delegation" key pair.

Following this, the selected participant applies the KEM protocol [16] to encapsulate their $sk'_i$ using the off-chain component's public key, $pk_{\text{off}}$. This results in an encrypted symmetric key and a ciphertext, which is then submitted to the contract. Detecting this event, the off-chain component decapsulates the received key using its private counterpart $sk_{\text{off}}$, thereby retrieving $sk'_i$.

With the delegated secret key $sk'_i$, the off-chain component can now generate the ring signature, effectively functioning as participant $i$. By utilizing $sk'_i$, the off-chain component can sign and simultaneously harness the set of public keys $R$ of the on-chain participants. This facilitates any entity to independently ascertain the ring signature against the aggregated public keys, which encompasses $pk'_i$, thereby validating the integrity and accuracy of the off-chain operations.

### 3.2 VRF Formal Instantiation:

Given our MPC-based Ring-LWE VRF system, we formalize the functions as:

1. **Key Generation:** Let $\mathcal{D}$ be the domain of all possible security parameters. Then, the function *Gen* is represented as:

   $$\text{Gen}: \mathcal{D} \to \mathcal{K} \times \mathcal{K}, \qquad where\ \text{Gen}(1^\lambda) = (PK, SK)$$

   for $\lambda \in \mathcal{D}$, and $(PK, SK)$ are the public and private key pairs for participants, respectively.

2. **Evaluate:** Let $\mathcal{S}, \mathcal{V}$ be the domains of all possible seeds and VRF system output of the tuple $(VRF_{output}, Proof_\pi)$. Then, the function *Eval* is represented as:

   $$\text{Eval}: \mathcal{S} \to \mathcal{V}, \qquad where\ \text{Eval}(Seed) = (VRF_{output}, \pi)$$

   for $Seed \in \mathcal{S}$, and $VRF_{output}$ is the generated output from the Ring-LWE encryption. More details are covered in Section 3.4.

3. **Verify:** Let $\mathcal{P}$ be the domain of all possible VRF proofs. Then, the function *Ver* is represented as:

   $$\text{Ver}: \mathcal{P} \to \{\texttt{TRUE}, \texttt{FALSE}\}, \qquad where\ \text{Ver}(\pi) = \texttt{TRUE}$$

   if and only if the NIZK proof $\pi$ is valid.

### 3.3 Core Computation Processes

The Algorithm 2 articulates the main computation process of our VRF system, based on a private DID-based MPC seed generation and the Ring-LWE encryption. For off-chain computation's integrity and signer's anonymity, we adapt the DID-based NTRU linkable ring signature system inspired by Tang et al. [15] along with the delegated key generation (DKG) mechanism with KEM [16].

This hybrid VRF system starts with the acquisition of a combined block hash derived from the three most recent blockchain blocks. Subsequent to this, the system instantiates DID keys, encompassing both the primary and the delegated keys. An essential component of our methodology is the delegation of keys, which provides the protocol with an encrypted secret key $sk_{enc}$ that encapsulates the system's inherent confidentiality needs.

The MPC process is used to derive the weighted sum of each commitment and share, based on each set of participants for a given round. This sum, in conjunction with the combined block hash, is used for the generation of a unique on-chain seed. Then, the seed is subsequently broadcasted on-chain, serving as the seed for the off-chain VRF cryptographic processes that ensue.

In the off-chain domain, the system delves into the core RLWE encryption, deriving cipher texts $(c1, c2)$. These cipher texts are then hashed to obtain 256-bit $VRF_{\text{output}}$ via the quantum-resistant keccak256 function, which encapsulates the randomness generated by the system. Using the off-chain secret key $sk_{off}$ and the previously delegated encrypted key $sk_{enc}$, the system can decrypt to obtain the desired $sk'_i$. This key is pivotal for the subsequent generation of the DID-based linkable ring signature on NTRU lattice [15], for our system's assurance of authenticity and non-repudiation.

The on-chain verification of the VRF's output leverages the VRF proof, ensuring that the output is both verifiable and random, without revealing the intricate details of the participants or the detailed computations.

In essence, our VRF algorithm embodies a harmonious blend of on-chain transparency and off-chain cryptographic rigor, ensuring verifiable randomness with robust quantum security strengths.

---

**Algorithm 3 Setup for DID-based NTRU Linkable Ring Signature Scheme**

1: **procedure** SETUP($1^\lambda, 1^N$)
**Require:** Security parameter $\lambda$, number of ring members $N$
**Ensure:** Public parameters $PP$, system master private key $MSK$
2:  Choose $k > 0$ s.t. $n = 2^k$
3:  Choose prime $q \equiv 1 \mod 2n$
4:  Calculate parameters $s$ and $\sigma$
5:  Set polynomial ring $R_q = (Z_q[x]/(x^n + 1))$
6:  Obtain $PP$ and $MSK$:
  – The $TrapGen_{NTRU}$(q,n,s) in [12] generates $h \in R_q$ and short basis $B$.
  – Select hash functions ($H_1 : \{0,1\}^* \to Z_q^n$ and $H_2 : \{0,1\}^* \to \{0,1\}^n$)
  – $MSK = B$; $MPK = h$
7:  Output public parameters $PP = (h, H_1, H_2)$ and keep $MSK = B$ secret
8: **end procedure**

---

### 3.4 DID-based Linkable Ring Signature on NTRU lattice with VRF Components

Our basic DID-based linkable ring signature scheme on NTRU lattice adapts the framework of Tang et al. [15]. The linkable ring signature ensures signer ambiguity, permitting verifiers to recognize a ring participant's signature without identifying the specific signer but with likability ensured to prevent malicious double signing. This methodology offers MPC anonymity against

**Algorithm 4** DID-based Linkable Ring Signature Scheme on NTRU lattice

1: **procedure** KeyGen(PP, $DID_i$, MSK)
**Require:** Public parameters $PP$, user's identity $DID_i$, system master private key $MSK = B$
**Ensure:** Pair of public/private key $(pk_i, sk_i)$
2:     Calculate the public key: $pk_i = t_{i,1} = H_1(DID_i) \in Z_q^n$
3:     Use CGS sampling algorithm [12]:
4:     Generate $(s_1, s_2) = (t_{i,1}, 0) - CGS(B, \sigma, (t_{i,1}, 0))$ such that $s_1 + s_2 \cdot h = t_{i,1}$
5:     Randomly choose polynomial vectors $s_1^*, s_2^* \leftarrow D_\sigma^n$
6:     Return user's public key $pk_i$ and private key $sk_i = (s_1, s_2, s_1^*, s_2^*)$
7: **end procedure**
8: **procedure** NTRU_LinkableRingSign(PP, R, m, $sk_k$)
**Require:** Public parameters $PP$, ring user identity set $R = \{DID_1, DID_2, \ldots, DID_N\}$, message $m \in \{0,1\}^*$, private key $sk_k = (s_1, s_2, s_1^*, s_2^*)$ for user $DID_k \in R$.
9:     Calculate $I = t_{k,1} + t_{k,2} \in R_q$, with $t_{k,2} = s_1^* + s_2^* \cdot h \in R_q$.
10:     **for** $i = 1$ to $N$ **do**
11:         Randomly select $y_{i,1}, y_{i,2} \leftarrow D_\sigma^n$, with corresponding vectors for short polynomials in $R_q$.
12:     **end for**
13:     Compute $v = H_2\left(\sum_{i=1}^N y_{i,1} + y_{i,2} \cdot h, R, m, I\right)$.
14:     **for** $i = 1$ to $N$ **do**
15:         **if** $i \neq k$ **then**
16:             Set $z_{i,1} = y_{i,1}, z_{i,2} = y_{i,2}$.
17:         **else**
18:             Compute $z_i = \begin{pmatrix} s_1 + s_1^* \cdot v + y_{i,1} \\ s_2 + s_2^* \cdot v + y_{i,2} \end{pmatrix}$.
19:         **end if**
20:     **end for**
21:     Output signature $\sigma_R(m) = \left((\mathbf{z}_{i,1}, \mathbf{z}_{i,2})_{1 \leq i \leq N}, v, I\right)$
22: **end procedure**
23: **procedure** VerifySign(PP, R, m, $\sigma_R(m)$)
24:     **for** $i = 1$ to $N$ **do**
25:         **if** NOT $(\|z_{i,1}\| \leq 2\sigma\sqrt{n}$ AND $\|z_{i,2}\| \leq 2\sigma\sqrt{n})$ **then return** "Invalid"
26:         **end if**
27:     **end for**
28:     **if** $H_2(\sum_{i=1}^N z_{i,1} + z_{i,2} \cdot h - I \cdot v, R, m, I) = v$ **then return** "Valid"
29:     **else return** "Invalid"
30:     **end if**
31: **end procedure**
32: **procedure** LinkSign($\sigma_R(m_1), \sigma_R(m_2)$)
33:     Input the two signatures $\sigma_R(m_1)$ and $\sigma_R(m_2)$
34:     Verify if $I(1) = I(2)$:
35:     **if** $I(1) = I(2)$ **then**
36:         Output "Link"
37:     **else**
38:         Output "Unlink"
39:     **end if**
40: **end procedure**

passive adversaries and is adaptively secure against chosen message attacks within the random oracle model. The basic setup is given in Algorithm 3 and the fully detailed procedures consist in Algorithm 4.

### 3.5 Delegated Key Generation (DKG)

We introduce a secure process to delegate the capability for an off-chain component to generate a DID-based linkable ring signature on a NTRU lattice [15] on behalf of an on-chain participant, without directly revealing any participant's identity.

---

**Algorithm 5** Delegated Key Generation (DKG) with Off-chain Signature Generation
---
1: **procedure** GENERATEKEYS(∅)
2:     $(sk_i, pk_i) \leftarrow$ KEYGEN() in Algorithm 4
3:     $(sk'_i, pk'_i) \leftarrow$ KEYGEN() in Algorithm 4
4:     **return** $(sk_i, pk_i, sk'_i, pk'_i)$
5: **end procedure**
6: **procedure** DELEGATEKEY($sk'_i, pk_{off}$)     ▷ $pk_{off}$ is the public key of the off-chain component
7:     $params \leftarrow Setup(1^\lambda)$
8:     $(c, K) \leftarrow Encaps(params, pk_{off})$     ▷ DKG Key Encapsulation
9:     $encryptedKey \leftarrow$ ENCRYPT($K, sk'_i$)
10:     **return** ($encryptedKey\ sk_{enc}, c, params, PP$)     ▷ PP from Algorithm 4
11: **end procedure**
12: **procedure** OFFCHAINSIGN($sk_{off}, sk_{enc}, c, params, PP, Seed, VRF_{output}, R$)
                                                                                                 ▷ $R$ is the set of DID public keys
13:     $K \leftarrow Decaps(params, c, sk_{off})$     ▷ DKG Key Decapsulation
14:     $sk'_i \leftarrow$ DECRYPT($K, sk_{enc}$)
15:     $m \leftarrow (Seed, VRF_{output})$
16:     $\sigma \leftarrow$ NTRU_LINKABLERINGSIGN($PP, R, m, sk'_i$)
17:     $\pi \leftarrow (VRF_{output}, Seed, \sigma)$
18:     **return** $\pi$
19: **end procedure**

---

**Protocol**

i. Every participant $i$ from the ring, generates two DID [14] key pairs: $(sk_i, pk_i)$ and $(sk'_i, pk'_i)$, as shown in Algorithm 5. The former is intended for on-chain operations while the latter acts as a "delegation" key pair.
ii. Participants commit both DID $pk_i$ and $pk'_i$ to the smart contract.
iii. For the MPC-based seed generation, participants utilize DID $sk_i$.
iv. Subsequent to MPC-based seed derivation, a chosen participant derives a KEM key $K$ using the public key of the off-chain component, denoted as DID $pk_{\text{off}}$. Using $K$, the participant encrypts them DID $sk'_i$ and the encrypted key is then submitted to the contract.
v. The off-chain component, upon detecting this event, decrypts the received value using its private key DID $sk_{\text{off}}$ and consequently obtains DID $sk'_i$. This key is then used for generating the ring signature.

**NTRU Linkable Ring Signature with Delegated Key Protocol** The off-chain component, with the delegated DID secret key $sk'_i$, can now perform the DID-based NTRU linkable ring signature generation, effectively acting as participant $i$, as shown in Algorithm 5.

i. Utilizing $sk'_i$, the off-chain component signs and simultaneously uses the consolidated public keys of the on-chain participants (inclusive of $pk'_i$).

ii. The non-interactive zero-knowledge (NIZK) proof, ensuring the signature's validity, is obtained via NTRU linkable ring signature.

**Verification** External verifiers can independently verify the NTRU linkable ring signature's correctness by using the collective DID public keys ($R$) in Algorithm 4, ensuring the correctness and integrity of the off-chain computation.

---

**Algorithm 6** Off-chain Blockchain Listener for Ring-LWE Computation
---
1: **function** RLWE_PROCESSING($seed$)
2:    **comment:** Assuming LWE takes seed value as a command-line argument
3:    $result \leftarrow$ subprocess.run(["./LWE", str($seed$)])
4:    **if** $result$.returncode $\neq 0$ **then**
5:       **raise Exception**(f'RLWE Encryption failed with error: result.stderr")
6:    **end if**
7:    $c1, c2 \leftarrow result.stdout$.strip()()
8:    **return** Ciphertexts $c1, c2$
9: **end function**

10: **function** HANDLE_EVENT($event$)
11:    $onchainSeed \leftarrow event['args']['onchainSeed']$
12:    $c1, c2 \leftarrow rlwe\_processing(onchainSeed)$
13:    $tx\_hash \leftarrow contract.functions.submitRLWEResult$(c1, c2).transact()
14:    $tx\_receipt \leftarrow w3.eth.waitForTransactionReceipt(tx\_hash)$
15:    **if** $tx\_receipt.status == 1$ **then**
16:       **print**("Result and proof submitted successfully!")
17:    **else**
18:       **print**("Submission failed.")
19:    **end if**
20: **end function**

21: **function** BLOCKCHAIN_LISTENER
22:    $w3 \leftarrow$ Web3(Web3.HTTPProvider('http://localhost:8545'))
23:    **comment:** Assuming contract ABI and address are available
24:    $contract \leftarrow w3.eth$.contract(address=contract_address, abi=contract_abi)
25:    $event\_filter \leftarrow$ contract.events.OnchainMpcSeedReady.$createFilter(fromBlock = 'latest')$
26:    **while** True **do**
27:       **for all** $event$ in $event\_filter$.get_new_entries() **do** handle_event($event$)
28:       **end for**
29:    **end while**
30: **end function**

---

### 3.6 Off-chain Ring-LWE Computation

The off-chain Ring-LWE Computation is implemented via a blockchain event listener tailored for the Ethereum platform. For testing purpose of deployment, we use Ganache [1] instead. At a high level, the listener waits for the `OnchainMpcSeedReady` event, and upon its detection, invokes the Ring-LWE encryption procedure. The main components and their sequential behaviors are as follows:

1. **RLWE Processing (`rlwe_processing` function):** This function is responsible for the execution of a external executable (LWE), which handles the Ring-LWE encryption process. Given a seed value (`onchainSeed`), the function invokes the executable LWE and captures its output.

---
[1] https://trufflesuite.com/ganache/

- Error checking is implemented to ensure the encryption process succeeds without hitches.
2. **Event Handling (`handle_event` function):** Once an event is detected, this function manages the subsequent steps. The main logic involves:
   - Extracting the `onchainSeed` argument from the event.
   - Using the `onchainSeed` as a seed for the RLWE encryption.
   - Transmitting the encrypted result (i.e., `c1` and `c2`) back to the Ethereum contract.
   - Monitoring the transaction receipt to confirm if the encrypted result submission was successful.
3. **Blockchain Listener (`blockchain_listener` function):** This is the core function that establishes a connection to the Ethereum network via the Web3 Python package. The logic flow includes:
   - Initializing a Web3 connection to a local Ethereum node.
   - Setting up the smart contract context using its ABI (Application Binary Interface) and address.
   - Implementing a continuous listening loop that actively monitors the `OnchainMpcSeedReady` event.
   - For each detected event, the `handle_event` function is invoked.

This listener, thus, seamlessly bridges on-chain events with off-chain Ring-LWE encryption processes, ensuring that encrypted results are promptly returned to the smart contract upon the trigger of specific events.

**Algorithm 7** RLWE_enc2 Encryption Algorithm
---
1: **procedure** RLWE_ENC2(a, c1, c2, m, p)
2:     encoded_m $\leftarrow m \times \frac{Q}{2}$
3:     $e_1, e_2, e_3 \leftarrow$ knuth_yao2() $\times 3$
4:     $e_3 \leftarrow e_3 +$ encoded_m
5:     $e_1, e_2, e_3 \leftarrow$ fwd_ntt2($e_1$), fwd_ntt2($e_2$), fwd_ntt2($e_3$)
6:     $c_1 \leftarrow e_2 + a \times e_1$
7:     $c_2 \leftarrow e_3 + p \times e_1$
8:     $c_1, c_2 \leftarrow$ rearrange2($c_1$), rearrange2($c_2$)
9: **end procedure**

### 3.7 Ring-LWE Encryption Function: `RLWE_enc2()`

The `RLWE_enc2()` function in Algorithm 7, based on the work [17], defines the encryption operation based on the Ring Learning With Errors problem. Given a message polynomial $m(x) \in R$, a public key polynomial $a(x) \in R$, and a public value polynomial $p(x) \in R$, where $R$ represents the ring of polynomials and NTT denotes the Number Theoretic Transform(NTT) [42], the RLWE encryption process is defined as follows:

1. Encode the message by multiplying it by $\frac{Q}{2}$ where $Q$ is a system parameter. Let encoded_m$(x) = m(x) \cdot \frac{Q}{2}$.
2. Randomly sample three error polynomials $e_1(x)$, $e_2(x)$, and $e_3(x)$ from some error distribution.
3. Update $e_3(x)$ as $e_3(x) = e_3(x) +$ encoded_m$(x)$.
4. Transform $e_1(x)$, $e_2(x)$, and $e_3(x)$ into the NTT domain.
5. Compute the ciphertext polynomials:
$$c_1(x) = e_2(x) + a(x) \cdot e_1(x)$$
$$c_2(x) = e_3(x) + p(x) \cdot e_1(x)$$
6. Rearrange the coefficients of $c_1(x)$ and $c_2(x)$ for transmission.

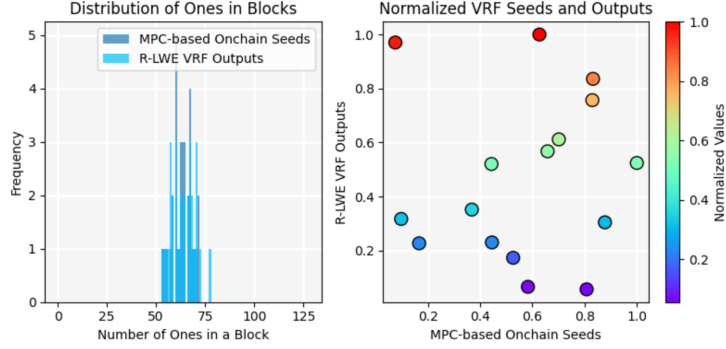

**Fig. 2.** The distribution of ones in 128-bit blocks and scattered distribution of VRF seeds and output. The average ratios of ones were observed as 0.4912 and 0.4934 in the MPC-based seeds and Ring-LWE VRF output, respectively.

## 4  Security and Privacy

### 4.1  Assumptions

Let $\lambda$ be the security parameter and $\epsilon(\lambda)$ be the negligible function in terms of the $\lambda$. When the setup phase of the key generation of our VRF system, *Gen*($1^\lambda$), is run with input $1^\lambda$, it outputs DID public and private keys for each participant. $Pr[\cdot]$ is the probability of an event.

- RING-LWE ASSUMPTION (QUANTUM SECURE): Given that quantum computers have efficient algorithms for discrete logarithm and integer factorization problems, but not (currently) for the Ring-LWE problem, we assume that it's computationally infeasible for a polynomial-time quantum adversary to solve the Ring Learning With Errors problem.
- QUANTUM RANDOM ORACLE MODEL (QROM): The hash function such as `keccak256()` is known to be quantum-resistant and behaves as a quantum random oracle, i.e., even quantum queries to the hash function cannot predict its outputs for new inputs.
- PRIVATE RING SIGNATURE UNFORGEABILITY: A polynomial-time adversary cannot forge a valid unique ring signature without knowledge of the private key.
- MPC SECURITY: It's computationally infeasible for any adversary to predict or influence the final MPC output unless it controls more than a threshold of the participants.

### 4.2  Formal VRF Requirements

**Uniqueness**  For a VRF system, uniqueness implies that for any given input, the function consistently yields the same output. That is, for any input message $m$, there exists a unique VRF output.

**Proof**: Let the function $\Phi$ represent our VRF system. Given an input $m$ (the MPC-based seed in our case), and the DID key pair ($DID_{sk}, DID_{pk}$) used for the VRF system: $\forall m_i, m_j$, where $i \neq j$,

$$\exists! \text{ VRF}_{\text{output}} \text{ s.t. } \Phi(m_i, DID_{sk}) = \Phi(m_j, DID_{sk}) = \text{VRF}_{\text{output}}$$

**Verifiability**  For a VRF system, verifiability ensures that if a prover produces an output $y$ and a proof $\pi$ for an input $x$, then a verifier can check the proof $\pi$ against the public key and confirm that $y$ is the correct VRF output for $x$.

**Proof**: Given the Input $m$ (the MPC-based seed), VRF output $y$, Proof $\pi$, DID Secret key $DID_{sk}$, Public key $DID_{pk}$.

If the prover produces $y$ and $\pi$ using $m$ and $DID_{sk}$:

$$y, \pi = \Phi(m, DID_{sk})$$

Then, the verifier, using $DID_{pk}$, can verify that $y$ is the correct VRF output for $m$:

$$\forall m, y, DID_{pk}, \pi, \ \Phi(m, DID_{pk}, \pi) \Rightarrow y = \text{VRF}_{\text{output}}(m)$$

This signifies that if the verifier utilizes the public key and proof to verify the input, they will consistently acquire the correct VRF output for that input.

**Pseudo-Randomness** For a VRF system, pseudo-randomness ensures that the output appears random and unpredictable. Given a VRF output $y$ for an input $x$, one cannot distinguish $y$ from a random value.

**Proof**: Given an Input $m$, VRF output $y$, Proof $\pi$, Adversary $\mathcal{A}$ trying to distinguish $y$ from a random value, without queries to the VRF output $y$, the advantage Adv of $\mathcal{A}$ in distinguishing $y$ from a random value is negligible:

$$\forall m, y, \text{ without queries to y, } \text{Adv}(\mathcal{A}(y)) \leq \epsilon(\lambda)$$

We've demonstrated that the MPC-based Ring-LWE VRF system integrated with the DID-based linkable ring signature on NTRU lattice satisfies the three pivotal properties of a VRF: uniqueness, verifiability, and randomness. These properties, complemented by the security guarantees from our previous discussions, affirm that the VRF system is robust and secure.

## 4.3 MPC-based Seed Integrity

The integrity of the MPC-based seed relies on the commitments made by participants. Under the assumption of a random oracle model for the hash function, the probability that an adversary can produce a commitment for a value without knowing that value is negligible. Formally:

$$\Pr[Seed' \leftarrow \mathcal{A}(\text{Commitments}) : Seed' = Seed] \leq \epsilon(\lambda)$$

## 4.4 Unforgeability under Chosen Message Attack

For all messages $m_1, m_2, \ldots, m_k$ chosen adaptively by $\mathcal{A}$, where signatures $\sigma_1, \sigma_2, \ldots, \sigma_k$ of VRF outputs are produced, the probability that $\mathcal{A}$ produces a new valid signature $\sigma^*$ for a new message $m^*$ without knowledge of the DID private key that signed the DID-based linkable ring signature on NTRU lattice is negligible.

$$\Pr[\sigma^* \leftarrow \mathcal{A}(m_1, \sigma_1, \ldots, m_k, \sigma_k) : \Phi(m^*, \sigma^*) = \text{True} \wedge m^* \notin \{m_1, \ldots, m_k\}] \leq \epsilon(\lambda)$$

## 4.5 Post-Quantum Security

**Definitions**

- Let $RLWE_{q,\chi}$ be the Ring-LWE problem with modulus $q$ and error distribution $\chi$.
- $\mathcal{A}_{RLWE}$ is a polynomial-time adversary $\mathcal{A}$ trying to solve the Ring-LWE problem.
- $\mathcal{A}_{SVP}$ is an adversary trying to solve the approximate SVP in ideal lattices.
- $\alpha$ is the approximation factor for the SVP problem.

**The Learning With Errors problem over Rings** Given a random polynomial $a$ from a ring $R_q$ and a "noisy" product $b = (a \times s) + e \mod q$ where $s$ is a secret polynomial and $e$ is an error polynomial drawn from $\chi$, the goal is to recover $s$ or distinguish $b$ from a random polynomial.

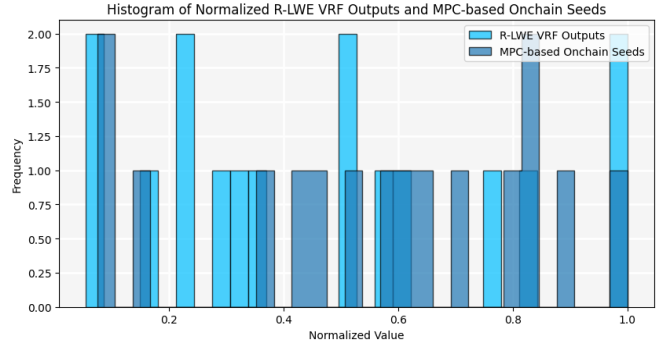

**Fig. 3.** Uniformly Distributed VRF outputs. The normalized values were derived from the raw 256-bit values.

**Security Proof for RLWE_enc2**

i. RING-LWE ASSUMPTION: It's computationally infeasible for a polynomial-time quantum or classical adversary to solve the Ring-LWE problem or distinguish between a valid Ring-LWE sample and a random one. ∀ Ring-LWE samples $s$,

$$Pr[s' \leftarrow \mathcal{A}_{RLWE}(a,b) : s' = s] \leq \epsilon(\lambda)$$

ii. REDUCTION TO SVP: If there exists a polynomial-time algorithm $\mathcal{A}_{RLWE}$ that can solve the $RLWE_{q,\chi}$ problem, then there exists an algorithm $\mathcal{A}_{SVP}$ that can solve the $\alpha$-approximate SVP in ideal lattices in polynomial time. ∀ lattices derived from Ring-LWE samples,

$$Pr[v' \leftarrow \mathcal{A}_{SVP}(Lattice) : ||v'|| \leq \alpha \times ||v_{shortest}||]$$

$$\geq Pr[s' \leftarrow \mathcal{A}_{RLWE}(a,b) : s' = s]$$

Here, $v_{shortest}$ is the shortest non-zero vector in the lattice. The $\alpha$-approximate SVP requires finding a vector whose length is within $\alpha$ times the shortest vector.

iii. POST-QUANTUM SECURITY: Given that SVP in ideal lattices is believed to be hard for quantum computers (there's no known polynomial-time quantum algorithm for this problem), the security of Ring-LWE and, in turn, `RLWE_enc2()` remains even in the presence of quantum adversaries. ∀ quantum adversary queries to $RLWE_{q,\chi}$,

$$Pr[s' \leftarrow \mathcal{A}_{RLWE}(QuantumQueries) : s' = s] \leq \epsilon(\lambda)$$

Overall, the security of the `RLWE_enc2()` function, as used in our VRF system, hinges upon the hardness of the Ring-LWE problem, which can be reduced to the hardness of the SVP in ideal lattices. This provides assurance of the post-quantum security of the function. The formal security for NTRU signature is referred to [12, Section 4] and [43] for lattice attacks and randomness properties of NTRU.

### 4.6 DKG Security

- **Confidentiality**: Employing asymmetric encryption ensures that $sk'_i$ remains confidential on-chain. Only the off-chain component, possessing $sk_{\text{off}}$, can decrypt this.
- **Integrity**: The ring signature confirms the integrity of the computation executed by the off-chain component.
- **Redundancy**: For backup, multiple participants might delegate their keys. This allows the off-chain component to choose from any of the provided keys, should one be unavailable.

– **Revocation**: A key pair update mechanism allows any participant to revoke or replace their delegation key pair, if they suspect potential misuse.
– **Non-repudiation**: The utilization of a specific delegated key for the ring signature holds the corresponding participant accountable, ensuring they cannot repudiate their involvement.

### 4.7 DID Privacy (GDPR Compliance)

Decentralized Identifiers (DID) [14] represents a paradigm shift in identity verification, emphasizing self-sovereign identity control and privacy. In the context of MPC-based VRF, where various nodes collaboratively generate a random number, DIDs offer an additional layer of obscurity. For instance, by obfuscating the linkage between real-world identities and their cryptographic counterparts, an adversary can't easily deduce a particular node's input value. Thus, integrating DIDs can make the VRF secure from such attempts, reinforcing the privacy of the system. Additionally, the interoperable nature of DIDs across various platforms and systems augments their utility, enabling a privacy-preserving environment for VRFs and beyond. With unique DIDs, the system ensures that participants can prove their identity without revealing any personal data. Our DID-based private ring signature scheme in Section 3.4 further ensures that even when a participant signs, their specific identity remains hidden among the members of the ring. DIDs, as a decentralized identity, provide users control over their identity without relying on centralized authorities. In this context, DIDs are used for verification rather than identification, ensuring participants' actions are verifiable without revealing their exact identities.

Given the DID-based ring signature and QROM, the ability of any adversary (including quantum adversaries) to link a signature to a specific DID or to single out any individual signer becomes negligible. Thus, the system satisfies GDPR[2] requirements in terms of unlinkability, inference protection, and prevention of singling-out. Formally: ∀ DID in the ring,

$$Pr[\text{DID}^* \leftarrow \mathcal{A}(\sigma) : \text{DID}^* \text{ is the actual signer}] \leq \epsilon(\lambda)$$

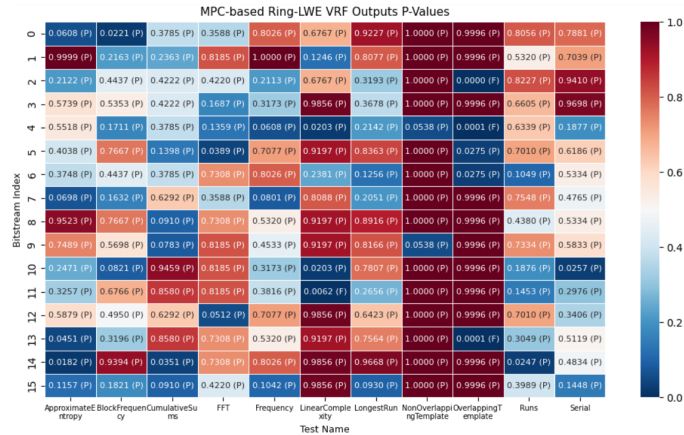

**Fig. 4.** *P*-value results from the NIST SP800-22 Test Suite

## 5 Complexity Analysis

In this section, we provide an in-depth complexity analysis of the proposed MPC-based hybrid VRF system using the Ring-LWE encryption and NTRU linkable ring signature. We break down the major components and evaluate both their time and space complexities, providing insights into the system's efficiency.

---
[2] https://gdpr-info.eu/recitals/no-26/

## 5.1 Temporal Complexity

- **MPC Seed Generation:** The primary operation is hashing, with a complexity of $\Theta(k)$ per hash, where $k$ denotes the fixed output size of the hash. Other basic arithmetic operations (addition, multiplication, modulo) are executed in constant time, $\Theta(1)$. However, these operations are repeated for each participant, leading to an overall complexity of $\Theta(n + k)$, where $n$ is the number of participants.
- **RLWE_enc2() Function:** This function involves polynomial multiplication, which predominantly determines its complexity. With $M$ being the polynomial degree, the complexity for polynomial multiplication is $\Theta(M \log M)$. Moreover, NTT and rearrangement operations, which grow with polynomial size, further contribute to the complexity, falling in the same ballpark of $\Theta(M \log M)$.
- **submitRLWEResult() Function:** The primary operation in this function is again hashing, which has a complexity of $\Theta(k)$ per hash operation.
- **NTRU Linkable Ring Signature Generation:** To analyze the temporal complexity of the DID-based Linkable Ring Signature Scheme on NTRU lattice algorithm, we'll break down each procedure and estimate the computational complexity of its steps in Algorithm 4.
    a. **KeyGen Procedure (see [12, Fig. 3])**
        - Calculating the public key using $H_1$ is likely $\Theta(N)$, assuming $H_1$ is a hash function with linear complexity and N represents the degree of the polynomials or the dimension of the NTRU lattice.
        - The CGS (Convolution Gaussian Sampler) algorithm is typically $\Theta(N \log N)$ due to the use of convolutions and FFT (Fast Fourier Transform) techniques.
        - Randomly choosing polynomial vectors is $\Theta(N)$.
        - Overall, the complexity of KeyGen is dominated by the CGS step, making it $\Theta(N \log N)$.
    b. **NTRU_LinkableRingSign Procedure**
        - Calculating $I$ involves basic arithmetic operations in $R_q$ which is $\Theta(N)$.
        - The loop for $i = 1$ to $n$ involves selecting polynomial vectors, each of which is $\Theta(N)$, resulting in $\Theta(Nn)$.
        - Computing $v$ with $H_2$ is $\Theta(Nn)$ assuming linear complexity for $H_2$.
        - The second loop also has $\Theta(Nn)$ complexity due to vector arithmetic in each iteration.
        - Thus, the overall complexity of NTRU_LinkableRingSign is $\Theta(Nn)$.
    c. **VerifySign Procedure**
        - The loop iterating over $n$ items and checking norms of vectors is $\Theta(Nn)$, assuming constant time for norm checking.
        - The final conditional statement involves a hash computation and arithmetic operations in $R_q$, which are $\Theta(Nn)$.
        - Hence, the VerifySign procedure is $\Theta(Nn)$.
    d. **LinkSign Procedure**
        - This procedure involves a simple comparison of the elements $I(1)$ and $I(2)$ from two signatures, which is $\Theta(1)$.
        - Therefore, the LinkSign procedure has constant complexity, $\Theta(1)$.

In conclusion, the most computationally intensive part of the NTRU linkable signature generation algorithm is the NTRU_LINKABLERINGSIGN and VERIFYSIGN procedures in Algorithm 4, both of which have a complexity of $\Theta(Nn)$. The KeyGen procedure, while involving a log factor due to the CGS algorithm, is generally less intensive for typical parameter sizes. The LinkSign procedure is the least intensive with constant complexity.

– **Aggregate Time Complexity:** The combined time complexity, accounting for all the aforementioned components, can be expressed as:

$$\Theta((N + 1)n + 2k + M \log M + \log p)$$

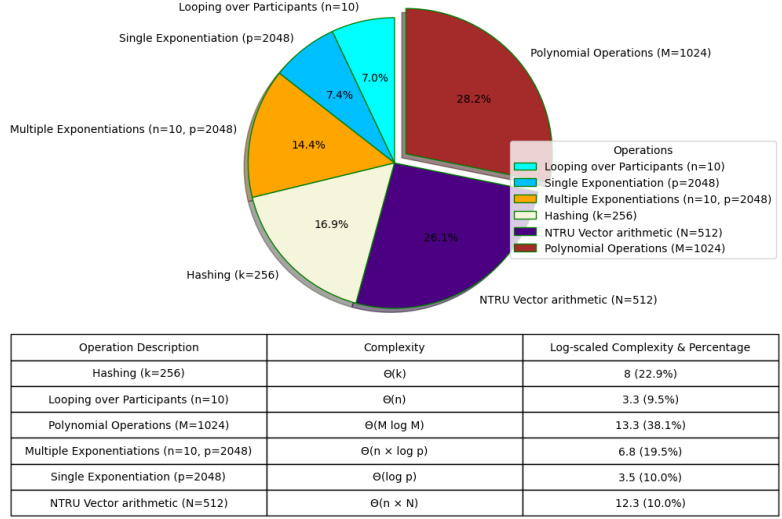

| Operation Description | Complexity | Log-scaled Complexity & Percentage |
|---|---|---|
| Hashing (k=256) | $\Theta(k)$ | 8 (22.9%) |
| Looping over Participants (n=10) | $\Theta(n)$ | 3.3 (9.5%) |
| Polynomial Operations (M=1024) | $\Theta(M \log M)$ | 13.3 (38.1%) |
| Multiple Exponentiations (n=10, p=2048) | $\Theta(n \times \log p)$ | 6.8 (19.5%) |
| Single Exponentiation (p=2048) | $\Theta(\log p)$ | 3.5 (10.0%) |
| NTRU Vector arithmetic (N=512) | $\Theta(n \times N)$ | 12.3 (10.0%) |

**Fig. 5.** Log-scaled Complexity Contributions in Hybrid VRF System.

## 5.2 Space Complexity

The space complexity of cryptographic systems is largely influenced by the data structures and variables used during computation.

– **MPC Seed Generation:** Given its hashing nature, it would require space proportional to the number of participants, i.e., $\Theta(n)$.
– **RLWE_enc2() Function:** Being polynomial-centric, it demands space proportional to the polynomial degree, leading to a space complexity of $\Theta(M)$.
– **submitRLWEResult() Function:** This function, being related to hashing and data verification, would require a space complexity of $\Theta(k)$, which is the output size of the hash.
– **NTRU Linkable Ring Signature Generation:** The space complexity analysis of the "DID-based Linkable Ring Signature Scheme on NTRU lattice" algorithm is as follows:
  1. **KeyGen Procedure:** Stores polynomials in $\mathbb{Z}_q^n$, leading to a space complexity of $\Theta(N)$.
  2. **NTRU_LinkableRingSign Procedure:** Requires space for $n$ identities and $2n$ vectors of size $N$ each, resulting in $\Theta(Nn)$.
  3. **VerifySign Procedure:** Operates on $n$ pairs of vectors, each of size $N$, contributing $\Theta(Nn)$ to the space complexity.
  4. **LinkSign Procedure:** Checks if two signatures are linked, depending on the size of the signatures, which is $\Theta(Nn)$.

  Thus, the dominant term in the space complexity for this algorithm is $\Theta(n)$, reflecting the storage requirements for the vectors and signature components.
– **Aggregate Space Complexity:** The combined space complexity, taking into account all the components mentioned, can be articulated as:

$$\Theta(n + k + M)$$

### 5.3 Efficiency Approximation

From our analysis, it's evident that the efficiency of the MPC-based Ring-LWE VRF system is predominantly influenced by the number of participants ($n$) and the polynomial degree ($M$). The constants $k$ and $p$ play a role but have a fixed impact due to their constant nature. For optimal system performance, it's imperative to manage the size of $n$ and $M$ in accordance with the computational capabilities of the deployment environment. This becomes particularly crucial if the system is intended for scenarios with frequent operations or in environments with constrained computational resources.

To approximate and highlight the dominant complexities derived in Section 5.1, 5.2, we first need to assign specific values to our variables: $k, n, M, p$, and $N$. These values are for approximation purposes, and in a real-world scenario, they might vary.

- $k$ (hash output size) = 256 (typical for cryptographic hashes)
- $n$ (number of participants) = 10
- $M$ (polynomial degree) = 1024 (example size for some RLWE operations)
- $p$ (modulus in group) = A large prime number with around 2048 bits
- $N$ (dimension of NTRU lattice) = 512 (example size for NTRU signature generation operations)

Given that $\log_2(2048)$ is approximately 11, we'll assume $\log p$ is around 11 for our computations.

With these values, we can compute the individual complexities:

- Hashing: $\Theta(k) = 256$
- Looping over participants: $\Theta(n) = 10$
- Polynomial operations: $\Theta(M \log M) = 1024 \times 10 = 10240$
- Exponentiations: $\Theta(n \times \log p) = 10 \times 11 = 110$
- Single exponentiation: $\Theta(\log p) = 11$
- NTRU Vector arithmetic: $\Theta(n \times N) = 5120$

Using a logarithmic function to scale the above complexities, we derive the following approximations according to the previous complexity analysis. Fig. 5 delineates these approximations, accompanied by a comprehensive summary table.

- Hashing: $\log_2(256) \approx 8$
- Looping over participants: $\log_2(10) \approx 3.3$
- Polynomial operations: $\log_2(10240) \approx 13.3$
- Exponentiations: $\log_2(110) \approx 6.8$
- Single exponentiation: $\log_2(11) \approx 3.5$
- NTRU Vector arithmetic: $\log_2(5120) \approx 12.3$

## 6 Evaluation and Deployment

### 6.1 Entropy Approximation

To estimate the entropy of our hybrid VRF system, we can consider the randomness generated by the VRF system to be a random variable $X$ that takes values in the range $[0, 2^{256} - 1]$. The entropy of $X$ can be estimated using the probability distribution function (PDF) of $X$. The PDF of $X$ can be derived as follows:

Table 1. Summary of NIST SP800-22 Test Suite Results

| Test Case Name | Total Tests | Average P-Values | Pass | Fail | Pass % |
|---|---|---|---|---|---|
| Approximate Entropy Test | 16 | 0.3929 | 16 | 0 | 100.0 |
| Frequency Test within a Block | 16 | 0.4246 | 16 | 0 | 100.0 |
| Cumulative Sums Test | 16 | 0.4107 | 16 | 0 | 100.0 |
| Discrete Fourier Transform (Spectral) Test | 16 | 0.5096 | 16 | 0 | 100.0 |
| Frequency (Monobit) Test | 16 | 0.4883 | 16 | 0 | 100.0 |
| Linear Complexity Test | 16 | 0.6370 | 15 | 1 | 93.75 |
| Test for the Longest Run of Ones in a Block | 16 | 0.5632 | 16 | 0 | 100.0 |
| Non-overlapping Template Matching Test | 16 | 0.8817 | 16 | 0 | 100.0 |
| Overlapping Template Matching Test | 16 | 0.6907 | 13 | 3 | 81.25 |
| Runs Test | 16 | 0.4968 | 16 | 0 | 100.0 |
| Serial Test | 16 | 0.5087 | 16 | 0 | 100.0 |
| **Total** | **176** | **0.5459** | **172** | **4** | **97.73** |

i. The hybrid VRF system generates a Ring-LWE encryption [17] from a seed value. The seed value is assumed to be uniformly distributed and independent of all other variables. Thus, the seed value has a PDF that is a uniform distribution over the range $[0, 2^{256} − 1]$.

ii. The Ring-LWE encryption and the NTRU linkable ring signature generation by the off-chain computation are deterministic functions of the seed value. Thus, their PDF are the same as the PDF of the seed value.

iii. The hybrid VRF system constructs the randomness using a distributed MPC [44] approach, where multiple participants collaborate to generate the randomness. The randomness is constructed as a weighted sum of commitments made by the participants. The weights used in the sum are determined by the shares of the participants.

iv. The commitments made by the participants are assumed to be uniformly distributed and independent of all other variables. Thus, each commitment has a PDF that is a uniform distribution over the range $[0, 2^{256} − 1]$.

v. The weights used in the sum are determined by the shares of the participants. The shares are assumed to be fixed and independent of all other variables. Thus, each share has a PDF that is a delta function at its value.

vi. The weights used in the sum are normalized to ensure that the resulting randomness is in the range $[0, 2^{256} − 1]$. Thus, the PDF of the resulting randomness is a truncated distribution of the sum of the commitments, where the truncation is at the value $2^{256} − 1$.

Let $C_i$ be the commitment made by participant $i$, and let $S_i$ be the share of participant $i$. Let $R$ be the resulting randomness generated by the hybrid VRF system. Let $n$ be the number of maximum participants of the VRF MPC smart contract. Then, the PDF of $R$ is given by:

$$PDF_R(r) = \frac{1}{Z} \int_0^{2^{256}-1} \left( \prod_{i=1}^{n} PDF_C(C_i) \right) \times \delta \left( r - \frac{\sum_{i=1}^{n} C_i \cdot S_i}{\sum_{j=1}^{n} S_j} \right) U(0, 2^{256} - 1)(r) dr \quad (3)$$

where $U(a, b)$ is the uniform distribution over the range $[a, b]$, and $Z$ is the normalization constant given by:

$$Z = \int_0^{2^{256}-1} \left( \prod_{i=1}^{n} PDF_C(C_i) \right) U(0, 2^{256} - 1) \times \left( \frac{\sum_{i=1}^{n} C_i \cdot S_i}{\sum_{j=1}^{n} S_j} \right) dr \quad (4)$$

To calculate the entropy of our hybrid VRF system, we need to first calculate the Shannon entropy [45] of the $PDF_R(r)$ formula. The Shannon entropy is given by the following formula:

$$H = -\int_0^{2^{256}-1} PDF_R(r) \log_2(PDF_R(r)) dr \quad (5)$$

Using the definition of $PDF_R(r)$, we have:

$$H = -\int_0^{2^{256}-1} \left[\frac{1}{Z}\int_0^{2^{256}-1}\left(\prod_{i=1}^n PDF_C(C_i)\right) \times \delta\left(r - \frac{\sum_{i=1}^n C_i \cdot S_i}{\sum_{j=1}^n S_j}\right) U(0, 2^{256}-1)(r)dr\right] \\ \times \log_2\left[\frac{1}{Z}\int_0^{2^{256}-1}\left(\prod_{i=1}^n PDF_C(C_i)\right) \times \delta\left(r - \frac{\sum_{i=1}^n C_i \cdot S_i}{\sum_{j=1}^n S_j}\right) U(0, 2^{256}-1)(r)dr\right] dr \quad (6)$$

This is a complex expression due to the nested integrals and delta function. Now, making use of the properties of the delta function:

$$\int_0^{2^{256}-1} f(r)\delta(r-a)dr = f(a) \quad (7)$$

We can simplify our expression for entropy. Additionally, as the commitments $C_i$ are uniformly distributed over $[0, 2^{256} - 1]$, their PDF is:

$$PDF_C(C_i) = \frac{1}{2^{256}} \quad (8)$$

Substituting this in, the entropy formula simplifies further:

$$H = -\int_0^{2^{256}-1} \left(\frac{1}{Z}\left(\frac{1}{2^{256}}\right)^n U(0, 2^{256}-1)(r)\right) \times \log_2\left(\frac{1}{Z}\left(\frac{1}{2^{256}}\right)^n U(0, 2^{256}-1)(r)\right) dr \quad (9)$$

Given that $U(0, 2^{256} - 1)(r) = 1$ for $r$ in $[0, 2^{256} - 1]$, this can be further simplified to:

$$H = -\left(\frac{1}{Z}\left(\frac{1}{2^{256}}\right)^n\right) \log_2\left(\frac{1}{Z}\left(\frac{1}{2^{256}}\right)^n\right) \times 2^{256} \quad (10)$$

This is the final specific formula for the estimated entropy of $H$.

## 6.2 Randomness Evaluation

The NIST SP800-22 [18] test suite provides a comprehensive analysis of the randomness characteristics of binary sequences. For our MPC-based hybrid VRF system with RLWE encryption and NTRU linkable ring signature, we subjected its output data to the tests within this suite. Specifically, we've selected 11 of the 15 standard tests, excluding the 'Binary Matrix Rank Test', 'Maurer's Universal Statistical Test', 'Random Excursions Test', and 'Random Excursions Variant Test' due to incompatibility with our data patterns.

The overall performance, as evidenced by the summary Table 1 and Fig. 4, is highly commendable. All tests produced an average $p$-value uniformly distributed within the range $[0, 1]$ above the significance level of 0.01, with average values around the 0.5 mark. Such results denote excellent randomness, as a perfectly random sequence would yield an average $p$-value of 0.5. The 'NonOverlappingTemplate' test even achieved an outstanding average $p$-value of 0.881706.

Furthermore, out of the 176 individual tests performed across all categories, only 4 have failed, resulting in an overall pass rate of approximately 97.73%. It's worth noting the 'LinearComplexity' test exhibited a single failure, and the 'OverlappingTemplate' test recorded three. Yet, a $p$-value falling outside the expected threshold is not an critical indictment of the system's overall randomness, but rather an indication of certain statistical outlier instances in specific scenarios.

In essence, our MPC-based Ring-LWE VRF system with NTRU linkable ring signature exhibits reliable performance in terms of randomness, as demonstrated by the NIST SP800-22 test suite results. This affirms the system's robustness and suitability for applications demanding high-quality random sequences.

### 6.3 System Deployment

We used Ganache to deploy our hybrid VRF system based on Truffle and Remix environment, which is a personal blockchain that allows to test and deploy smart contracts on a local network without incurring the cost and time delay associated with deploying on the main Ethereum network. The unique contract address, comprised of 160 bits, is represented by the hexadecimal value "0xe256 1068 e91d 66db 8ecc 3c56 b695 03ce 1593 4f60". Concurrently, the contract creation transaction hash is given by "0xc9a9 39b8 286d 0e01 fcf8 edd3 b563 2389 0173 81b5 838f 65c2 f0f7 b27f a637 2654". The execution of the creation transaction necessitated the expenditure of 3,050,857 gas units. Fig. 3 provides a visual representation of the output VRF distribution, where the uniform distribution property is observed for both the MPC-based VRF on-chain seeds and the final R-LWE VRF outputs. Fig. 2 shows more characterized uniform distribution of ones in histogram along with scattered dispersion of the generated MPC-based seeds and Ring-LWE VRF outputs. Fig. To illustrate as an example, the 256-bit VRF output value was "0x2865 7095 a002 2a9a 64fe f449 29e1 1aee b56c 39cd bcf4 f14d f13c 1c5b f18b 6a5c" when the MPC-based on-chain seed was given as "0x5caa 7f6e 442b a853 8593 bb1f a85a 8894 5179 e846 15d5 f8cb 8ff6 a550 7091 3e59".

### 7 Conclusion

In response to the emerging threats posed by quantum computing on classical cryptographic protocols, this work introduces a novel and hybrid Verifiable Random Function (VRF) framework tailored for blockchain systems. By harnessing the post-quantum security of Ring-LWE encryption and linkable ring signature on NTRU lattice, we established a model that robustly generates pseudo-random sequences, maintaining security in the face of quantum challenges. Recognizing the computational intensity and the consequent on-chain gas costs due to the Ring-LWE encryption and NTRU linkable signature, a hybrid architecture was proposed. This architecture adaptively integrates the on-chain seed generation and the subsequent off-chain intensive computations for VRF evaluation and proof generation, with the validity proof through the DID-based linkable ring signature scheme on NTRU lattice and delegated key generation mechanism for off-chain computations. By incorporating multi-party computation (MPC) with blockchain-based decentralized identifiers (DID), our VRF system magnifies its collective randomness and fortifies its security with privacy. The subsequent evaluations underscore the significant security and privacy merits of our proposed model while presenting empirical evidence of its randomness through the NIST SP800-22 test suite. Our VRF model has demonstrated outstanding randomness performance, achieving a remarkable 97.73% overall pass rate on 11 standard tests and 0.5459 of average $p$-value for the total 176 tests. This attests that our VRF scheme is not only theoretically robust but also practically relevant for scenarios requiring verifiable randomness in blockchain environments.

### Acknowledgment

This research is supported by the Macao Polytechnic University research grant (Project code: RP/FCA-02/2022) and SUNY Korea with the National Research Foundation of Korea (NRF) grant funded by the Ministry of Science and ICT (MSIT), Korea (No. 2020R1F1A1A01070666).

## Authors

**Bong Gon Kim** holds a B.S. in Electrical Engineering and an M.S. in Computer Science from Yonsei University, Seoul. He is now a Ph.D. candidate at Stony Brook University's Computer Science Department. From 2003 to 2018, he was a Senior Engineer at Samsung Electronics, in mobile experience (MX) and S.LSI divisions contributing to Galaxy smart devices and Exynos platforms. His research focuses on DID, quantum computing, cryptography, and blockchain.

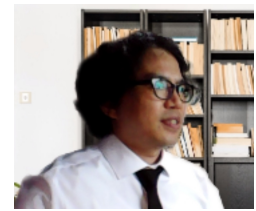

**Dennis Wong** earned his Ph.D. in Computer Science from the University of Guelph, mentored by Joe Sawada, after completing his undergraduate studies at the Chinese University of Hong Kong under Evangeline Young. His expertise spans combinatorics, algorithm design, graph theory, and sequence generation. Recently, his interests have extended to financial modeling and sports analytics.

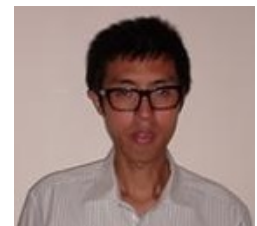

**Yoon Seok Yang** completed his B.S. and M.S. at Hanyang University, South Korea, followed by an M.S. from the University of California, Irvine, and a Ph.D. from Texas A&M University. His stint at LG Electronics as a Research Engineer led to significant contributions to Digital TV Labs. Now at SUNY Korea, his work spans neuromorphic computing, AI-based SoC design, machine learning signal processing, and edge computing solutions.

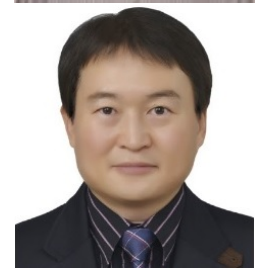